\documentclass[prb,twocolumn,showpacs,superscriptaddress,amsmath,amssymb]{revtex4}

\usepackage{graphicx}
\usepackage{dcolumn}
\usepackage{bm}
\usepackage[dvips]{color}
\usepackage{ulem}

\newcommand{\Li}{Li(Cu$_{1-x}$Zn$_{x}$)$_2$O$_2$}
\newcommand{\vect}[1]{\mathbf{#1}}
\newcommand{\figwidth}{0.98\columnwidth}

\newcommand{\vectmu}{\boldsymbol\mu}
\begin{document}

\title{ Magnetic structure of the frustrated $S=1/2$ chain magnet LiCu$_2$O$_2$ doped with nonmagnetic Zn}

\author{A.A. Bush}
\affiliation{Moscow State Technical University of Radioengineering, Electronics and Automation, 119454, Moscow, Russia}

\author {N. B\"{u}ttgen}
\affiliation{Center for Electronic Correlations and Magnetism EKM,
Experimentalphysik V, Universit\"{a}t Augsburg, D--86135 Augsburg, Germany }

\author {A.A. Gippius}
\affiliation{Center for Electronic Correlations and Magnetism EKM,
Experimentalphysik V, Universit\"{a}t Augsburg, D--86135 Augsburg, Germany }

\affiliation{Moscow State University, 119899, Moscow, Russia}

\author{V.N. Glazkov}
\affiliation{P. L. Kapitza Institute for Physical Problems RAS,
119334 Moscow, Russia}

\affiliation{Neutron Scattering and Magnetism, Laboratory for Solid
State Physics, ETH Z\"{u}rich, Switzerland,  8093 Z\"{u}rich,
Switzerland}

\author {W. Kraetschmer}
\affiliation{Center for Electronic Correlations and Magnetism EKM,
Experimentalphysik V, Universit\"{a}t Augsburg, D--86135 Augsburg, Germany }

\author{L.A. Prozorova}
\affiliation{P. L. Kapitza Institute for Physical Problems RAS,
119334 Moscow, Russia}

\author{L.E. Svistov}

\email{svistov@kapitza.ras.ru}

\affiliation{P. L. Kapitza Institute for Physical Problems RAS,
119334 Moscow, Russia}

\author{A.M. Vasiliev}
\affiliation{P. L. Kapitza Institute for Physical Problems RAS,
119334 Moscow, Russia}

\author{A. Zheludev}
\affiliation{Neutron Scattering and Magnetism, Laboratory for Solid
State Physics, ETH Z\"{u}rich, Switzerland,  8093 Z\"{u}rich,
Switzerland}

\author{A.M. Farutin}
\affiliation{Laboratoire Interdisciplinaire de Physique, CNRS et Universit\'e J.Fourier-Grenoble I, BP 87, 38402 Saint-Martin d'Her\`es, France}

\date{\today}

\begin{abstract}
We present the results of magnetization, ESR and NMR measurements on single
crystal samples of the frustrated $S=1/2$ chain cuprate LiCu$_2$O$_2$ doped
with nonmagnetic Zn$^{2+}$. As shown by the x-ray techniques the crystals of
\Li\ with $x<0.12$ are single-phase, whereas for higher Zn concentrations the
samples were polyphase. ESR spectra for all monophase samples ($0\leq x<0.12$)
can be explained within the model of a planar spin structure with a uniaxial
type anisotropy. The NMR spectra of the highly doped single crystal sample
Li(Cu$_{0.9}$Zn$_{0.1}$)$_2$O$_2$ can be described in the frame of a planar
spin-glass like magnetic structure with short-range spiral correlations in the
crystal ($ab$)-planes with strongest exchange bonds. The value of magnetic
moments of Cu$^{2+}$ ions in this structure is close to the value obtained for
undoped crystals: (0.8 $\pm$ 0.1) $\mu_B$.

\end{abstract}

\maketitle
\nopagebreak[4]

\section{Introduction}
LiCu$_2$O$_2$ is an example of a $S=1/2$ magnet with frustrated exchange
interactions. The magnetic structure of LiCu$_2$O$_2$ can be considered as a
system of coupled $S=1/2$ chains of magnetic Cu$^{2+}$ ions. Within the chains
the nearest spins interact ferromagnetically and the next-nearest neighbors
antiferromagnetically. Interest in magnetic systems with such a type of
frustration is stimulated by theoretical predictions of unusual magnetic phases
for uncoupled and weakly coupled (1D and quasi-1D) chain models (see for
example Refs.~\onlinecite{Hikihara_2008, Sudan_2009}). It was shown that a new
type of magnetic ordering can be realized for the  magnets with this particular
type of frustration. Such a type of magnetic order, with zero average magnetic
moment on each magnetic ion, but with long-range ordered correlations of spin
components of neighboring ions, is classified as spin-nematic
order.\cite{Andreev_1984} The possibility to find this new type of magnetic
order experimentally is a very attractive task. According to
Ref.~\onlinecite{Hikihara_2008}, for the 1D model with the intrachain exchange
constants of LiCu$_2$O$_2$  a chiral long-range order in the low-field range
and a quasi long-range ordered spin-density wave phase in higher applied
magnetic fields $H$ are expected. In the experiment the planar spiral spin
structure was observed at $T < T_N$ in the low-field range in
LiCu$_2$O$_2$.\cite{Masuda_2004, Gippius_2004} At higher fields $\mu_0H\approx
15$ T another magnetic phase, which is not identified yet, was
observed.\cite{Bush_2012} Probably, this phase is related to a spin-density
wave phase which is predicted for the 1D model. A spin-nematic phase for the
exchange integrals of LiCu$_2$O$_2$ is expected  in even higher magnetic fields
which are hardly accessible experimentally.

Introduction of non-magnetic impurities substituting for the magnetic ions
provides a powerful tool to fine tune properties of the bulk magnet and to
affect stability of different phases or even to stabilize new phases. The
effect of non-magnetic doping was studied in the collinearly ordered magnets
(e.g. Refs.~\onlinecite{Carretta_1997, Liu_2013}), in the quantum magnets (e.g.
Refs.~\onlinecite{tlcucl3-mg, dan2012}). Recent publications (e.g.
Refs.~\onlinecite{Papinutto_2005, Wollny_2011, Sen_2011, Sen_2012}) have
discussed the effect of non-magnetic impurities on the matrix with frustrated
interactions. Quasi-one-dimensional frustrated magnets are appealing objects
for the study of such doping effects: first, breaking up the spin-chain by an
impurity can lead to the formation of the lengthy multi-spin defect with
unusual properties.\cite{Zhitomirsky_13} Besides, in the presence of the strong
next-nearest neighbor interaction unaffected by doping, substitution of the
magnetic ion by non-magnetic ones results in the conservation of the magnetic
correlations along the chain with the defined spin spiral phaseshift at the
impurity location, while, due to the random location of the defects, weak
interchain interaction becomes frustrated.

The investigation of the zinc-doped crystals of  LiCu$_2$O$_2$ was reported
recently \cite{Hsu_2010} and new magnetic phases in this system were announced.
This report has stimulated the Electron Spin Resonance (ESR) and Nuclear
Magnetic Resonance (NMR) investigations  of the magnetic structure of
 Zn-doped LiCu$_2$O$_2$ single crystals reported here.
Our x-ray, ESR and NMR research provides no support for the existence of the
dimer phase at doping levels above $ x \approx 5\%$, suggested by Hsu et
al.\cite{Hsu_2010} On the contrary, we have found that at the doping up to 10\%
resonance response of the system remains qualitatively similar to that of the
pure sample and can be understood in terms of planar noncollinear structure.
The NMR study of the \Li\ single crystal with $x=0.1$ at $T<T_N$ shows
coexistence of static correlations corresponding to the magnetic structure of
the undoped material with a static random distribution of spin directions. The
temperature evolution of NMR spectra can be described by the development of the
interplane correlations with decrease of the temperature. For the temperatures
$T_{N}/2 < T < T_N$ the NMR spectra can be described in the model of
uncorrelated $(ab)$ spin planes.

The paper is built as follows. In the first section a short overview of the
literature data about the crystal and magnetic structures of the undoped
samples of LiCu$_2$O$_2$ is given.  In the second section of the paper the
details of crystal growth and sample characterization are described. In the
next two sections results of ESR and NMR experiments are presented.

\section{Crystallographic and magnetic structures of LiCu$_2$O$_2$}

LiCu$_2$O$_2$ crystallizes in an orthorhombic system (space group $Pnma$) with
unit cell parameters $a=5.73$ \AA, $b=2.86$ \AA \, and $c=12.42$
\AA.\cite{Berg_1991} The unit cell parameter $a$ is approximately twice the
unit cell parameter $b$. Consequently, the LiCu$_2$O$_2$ samples, as a rule,
are characterized by twinning due to the formation of crystallographic
domains rotated by $90^\circ$ around their common crystallographic $c$-axis.

 The unit cell of the LiCu$_2$O$_2$ crystal contains
four univalent nonmagnetic Cu$^+$ cations and four divalent Cu$^{2+}$ cations
with spin $S=1/2$. There are four crystallographic positions of the
magnetic Cu$^{2+}$ ions in the crystal unit cell of LiCu$_2$O$_2$,
conventionally denoted as $\alpha$, $\beta$, $\gamma$ and $\delta$.

\begin{figure}
\includegraphics[width=\figwidth,angle=0,clip]{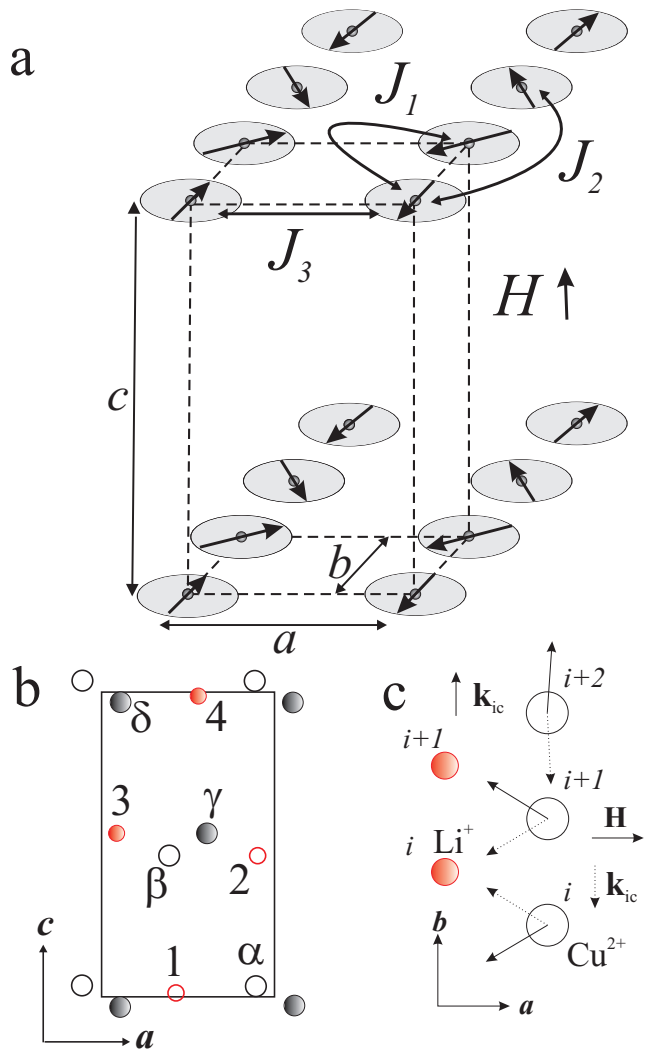}
\caption{a. (color online)  Schematic representation of the
arrangement of Cu$^{2+}$ moments in LiCu$_2$O$_2$. Only one of four Cu$^{2+}$
($\alpha$, $\beta$, $\gamma$, $\delta$) positions is shown. Arrows correspond
to the spin directions below $T_{c2}$ for $\vect{H}
\parallel \vect{c}$. $J_1, J_2$ and $J_3$ are the main exchange
integrals.\cite{Masuda_2005} b. Schematic arrangement of the Cu$^{2+}$ and
Li$^{+}$ ions in the projection of the crystal lattice onto the $ac$-plane
\cite{Svistov_2009}. Large circles indicate the Cu$^{2+}$ ions, and small
circles -- the Li$^{+}$ ions. Closed and open circles represent ions with the
coordinates equal to $0.25b$ and $0.75b$ along the $b$-axis, respectively. c.
Scetch of the spin configuration corresponding to the extreme values of dipolar
fields at Li site for two incommensurable domains with incommensurability
vector directed upchain (solid arrows) or downchain (dashed arrows). All spin
vectors and the external field are in plane of figure.} \label{fig:spins}
\end{figure}

 The Cu$^{2+}$ ions in  different
positions are weakly magnetically coupled and form four almost independent
systems of spin chains. Such chains formed by one of four kinds of Cu$^{2+}$
ions (e.g. the $\alpha$-position) and relevant exchange interactions\cite{Masuda_2005}
within the system are shown schematically in
Fig.~\ref{fig:spins}a.

The positions of Cu$^{2+}$ and Li$^+$ ions in the crystal cell are
given\cite{Berg_1991} by copper coordinates: (0.876; 0.75; 0.095), (0.376;
0.75; 0.405), (0.624; 0.25; 0.595), (0.124; 0.25; 0.905) and by lithium
coordinates: (0.376;0.75;0.068),
(0.876;0.75;0.432),(0.124;0.25;0.568),(0.624;0.25;0.932). Schematic arrangement
of the Cu$^{2+}$ and Li$^{+}$ ions in the projection of the crystal lattice
onto the $ac$-plane is given in Fig.~\ref{fig:spins}b.

The transition into the magnetically ordered state occurs via two stages at
$T_{c1}=24.6$~K and $T_{c2}=23.2$~K.\cite{Seki_2008} Neutron scattering and
NMR experiments have revealed that an incommensurate magnetic structure is
realized in the magnetically ordered state ($T<T_{c1}$).\cite{Masuda_2004,
Gippius_2004, Kobayashi_2009} The wave vector of the incommensurate magnetic
structure coincides with the chain direction (the $\vect{b}$-axis). The magnitude of
the propagation vector at $T<$ 17 K is almost temperature independent and is
equal to 0.827$\times$2$\pi$/$b$. The neutron scattering experiments have shown
that the neighboring magnetic moments along the $\vect{a}$-direction are antiparallel,
whereas those along the $\vect{c}$-direction are parallel. The investigation of the
spin-wave spectra by inelastic neutron scattering \cite{Masuda_2005}
shows that the incommensurate magnetic structure in LiCu$_2$O$_2$ is caused by
a competition between the ferromagnetic exchange interaction of the nearest-neighbor
magnetic ions in the chain $J_1=-7.00$~meV and the antiferromagnetic
interaction of the next-nearest neighbor ions in the chain $J_2=3.75$~meV. The
antiparallel orientation of the magnetic moments of Cu$^{2+}$ between
neighboring chains is caused by the strong antiferromagnetic interaction
$J_3=3.4 $~meV. The coupling of the Cu$^{2+}$ magnetic moments along the $\vect{c}$-direction
and the couplings between the magnetic ions in different crystallographic
positions $\alpha$, $\beta$, $\gamma$ or $\delta$ are much weaker.
\cite{Masuda_2005, Gippius_2004} Thus, the magnetic
structure of LiCu$_2$O$_2$ in the magnetically ordered phase can be considered
as quasi-two dimensional. The quasi-two-dimensional character of the magnetic
interactions in LiCu$_2$O$_2$ was also proven by resonant soft
X-ray magnetic scattering experiments.\cite{Rusydi_2008, Huang_2008}

The magnetic structure of LiCu$_2$O$_2$ at zero magnetic field was studied
using different experimental methods, but so far there is no generally
accepted model for it. For a review of proposed models see, for example,
Ref.~\onlinecite{Bush_2012}.

In magnetic fields above 3~T, the magnetic structure realized in LiCu$_2$O$_2$
is much clearer. ESR and NMR studies of LiCu$_2$O$_2$ in the low-temperature
magnetically ordered phase ($T<T_{c2}$) established that a planar spiral
magnetic structure is formed in this compound.\cite{Svistov_2009} The magnetic
moments located at the $\alpha$ ($\beta$, $\gamma$ or $\delta$) position of the
crystal unit cell with coordinates $x$, $y$, $z$ (measured along
 the $\vect{a}$, $\vect{b}$ and $\vect{c}$-axis of
the crystal, respectively) are defined as:

\begin{eqnarray}
\vectmu_\alpha=\mu \cdot \vect{l}_1 (-1)^{x/a} \cdot
\cos(k_{ic}\cdot y+\phi_\alpha)+ \nonumber\\
+\mu \cdot \vect{l}_2 (-1)^{x/a} \cdot \sin(k_{ic}\cdot
y+\phi_\alpha), \label{eqn:magnstr}
\end{eqnarray}

\noindent where $\vect{l}_1$ and $\vect{l}_2$ are the two mutually
perpendicular unit vectors, $\mathbf{k}_{ic}$ is the incommensurability vector
parallel to the chain direction ($\vect{b}$-axis), $\mu$ is the magnetic moment
of the Cu$^{2+}$ ion and $x$ is a multiple of the chain period $a$. The magetic
moment per copper ion at $T\lesssim 10$~K was evaluated as
$\mu=0.85$~$\mu_{B}$.\cite{Masuda_2004,Svistov_2009} The phases $\phi_\alpha$,
$\phi_\beta$, $\phi_\gamma$ and $\phi_\delta$ determine the mutual orientation
of the spins in the chains formed by the ions in the different crystallographic
positions. Their values are extracted from the NMR
data\cite{Svistov_2009, Svistov_2010}: $\phi_\alpha=0$; $\phi_\beta=\phi_\alpha + \pi/2$;
$\phi_\gamma=\pi-k_{ic}*b/2$; $\phi_\delta=\phi_\gamma+\pi/2$.

\begin{figure}
\includegraphics[width=\figwidth,angle=0,clip]{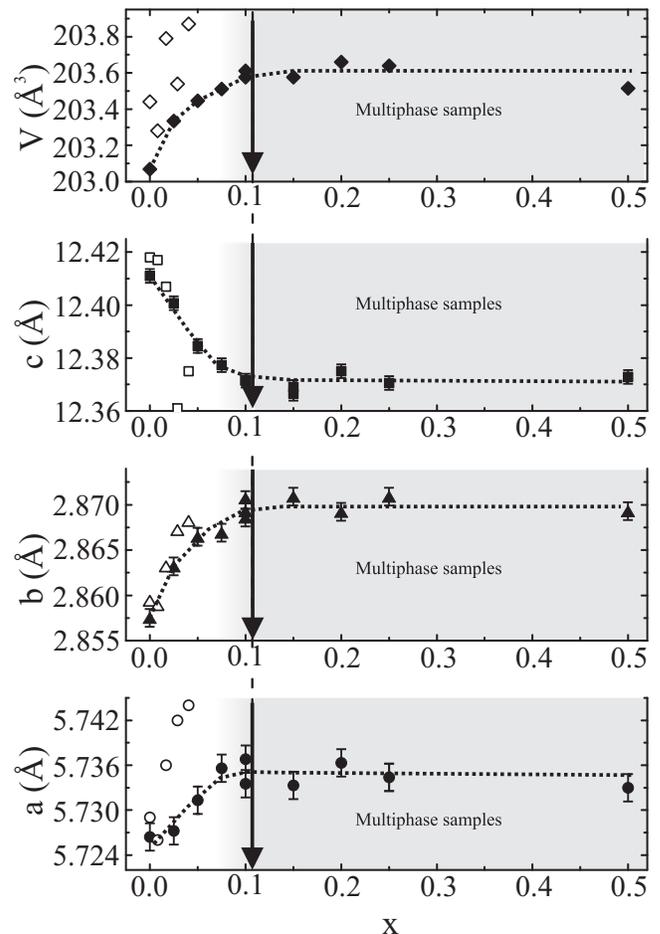}
\caption{The concentration dependence of the crystal unit cell volume (top panel)
and crystal unit cell parameters $a$,$b$,$c$ (lower panels) of single crystals of \Li.
$x$ $-$ concentration of zinc in initial charge. The solid symbols show the
results obtained on our samples discussed here. The open symbols show the data
taken from Ref.~\onlinecite{Hsu_2010}.} \label{fig:2}
\end{figure}

\begin{figure}
\includegraphics[width=\figwidth,angle=0,clip]{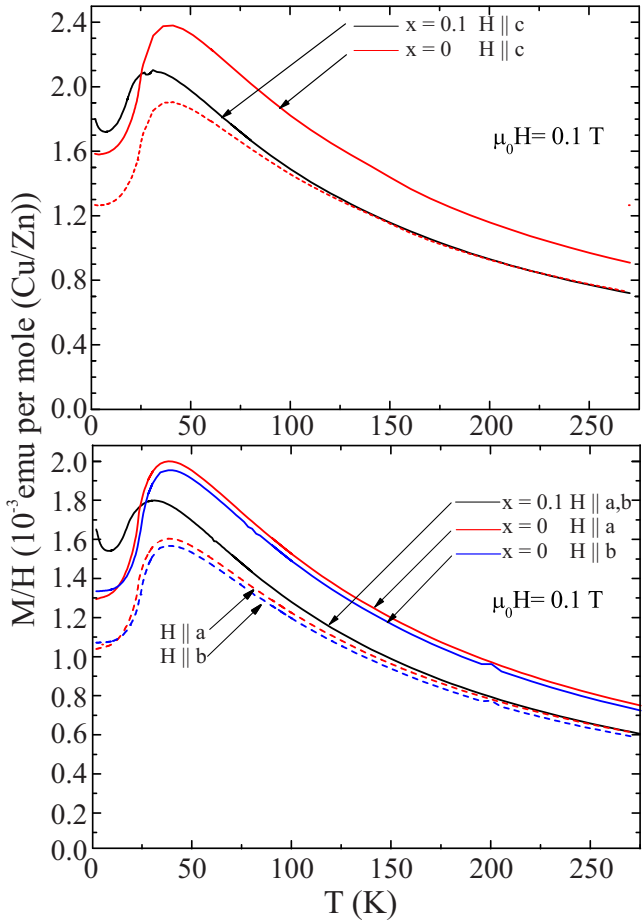}
\caption{(color online) Temperature dependences of the magnetic susceptibility
$M(T)/H$ for field directions $\vect{H}\parallel \vect{c}$ (upper panel) and
$\vect{H}\parallel\vect{a},\vect{b}$\ (lower panel) at $\mu_0 H=0.1$~T.
Susceptibility of the Zn-doped sample with $x=0.1$ was measured on the twinned
single crystal, data for the untwinned crystal with $x=0$ at $\vect{H}\parallel
\vect{a},\vect{b}$ and $\vect{c}$ are taken from Ref.~\onlinecite{Bush_2012}.
Dashed lines show the same $M(T)/H$  curves for undoped samples ($x=0$), scaled
on the $\vect{y}$-axis by a factor of $1-2x=0.8$.} \label{fig:M(T)}
\end{figure}

\begin{figure}
\includegraphics[width=\figwidth,angle=0,clip]{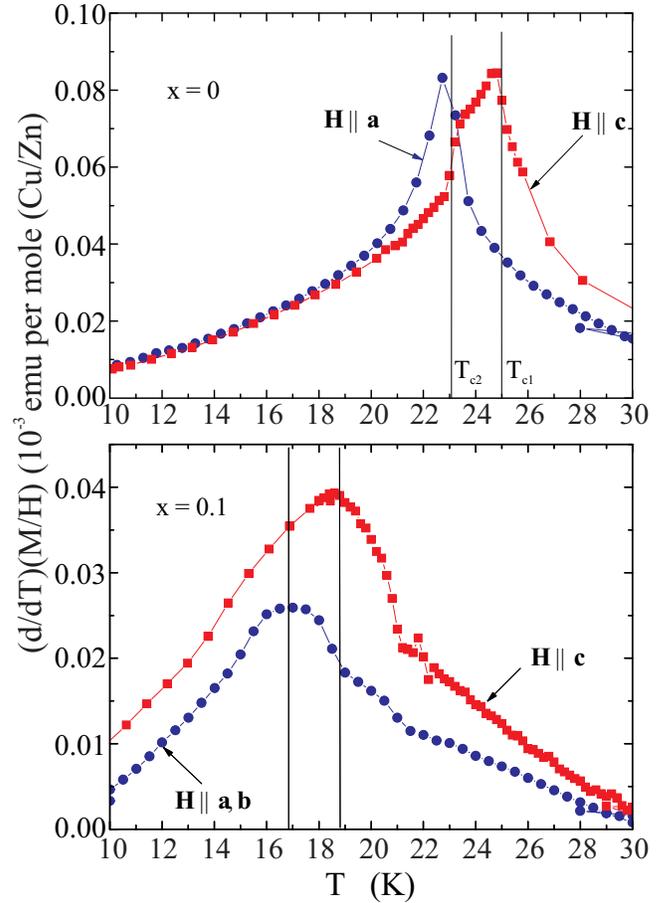}
\caption{(color online) Temperature dependences of the magnetization derivative
$dM/dT$ for field directions  $\vect{H}
\parallel \vect{a}$, $\vect{b}$ (circles) and $\vect{c}$ (squares) and
$\mu_0 H=0.1$~T for the undoped untwinned crystal $x=0$ (upper panel, taken
from Ref.~\onlinecite{Bush_2012}) and for the twinned doped single crystals of
\Li\ with $x=0.1$ (lower panel). Data are obtained from $M(T)/H$ at
$\mu_{0}H=$0.1~T.  } \label{fig:dmdt}
\end{figure}

\section{Sample preparation and experimental details}

Single crystals of \Li\ with the size of several cubic millimeters were
prepared by the ``solution in the melt'' method.\cite{Bush_2004}  Samples
were grown in air by a flux method in alundum crucibles. The mixtures of
analytical grade Li$_2$CO$_3$, CuO and ZnO of
Li$_2$CO$_3\cdot$4(1-x)CuO$\cdot$4xZnO compositions were melted at
1100~$^{\circ}$C, and then solidified by cooling to 930~$^{\circ}$C at a rate of
5.0~$^{\circ}$C/hour. Next, the crucible was withdrawn from the furnace and
placed on a massive copper plate to ensure rapid cooling to room temperature.
Quenching from $\approx$900~$^{\circ}$C was necessary, because  the single
phase LiCu$_2$O$_2$ single crystals decompose below this
temperature.\cite{Bush_2004}

The samples were shaped as a flat plates with the developed ($ab$)-plane.
The twinning structure of the samples was studied with optical polarization
microscopy.\cite{Bush_2012} In most cases, the samples were twinned with a
characteristic domain size of several microns. It was possible to select
samples without twinning structure for Zn concentrations up to $x=0.1$. The
absence of the twinning structure in the samples selected for the experiments
was confirmed by X-ray diffraction and ESR measurements.\cite{Svistov_2009,
Bush_2012} We have used samples without twinning for most of the experiments
described below.

The X-ray diffraction patterns were taken in a $\theta-2\theta$ geometry using
CuK$\alpha$ irradiation. The samples produced clearly distinguishable
diffraction patterns from the $(bc)$- and the $(ac)$-planes of the crystal. The
spectra obtained from twinned samples had a form of a superposition of these
spectra. The Zn doping of LiCu$_2$O$_2$ results in a monotonic change of the
cell parameters up to a concentration limit of $x_c \approx 0.12$. For samples
with zinc content of initial charge $x < 0.12$ no diffraction patterns due to
impurity phases were observed.  Additional x-ray diffraction patterns were
observed for the samples with $x\geq 0.12$, which can be ascribed to the
presence of inclusions of impurity phases. The patterns ascribed to the main
phase show that the cell parameters for higher Zn content do not change. The
zinc concentrations of \Li\ single crystals were measured by electron probe
micro analysis on the spectrometer ``Eagle II'' (``EDAX'', USA). The
concentration values of $x$ in our single-crystalline samples obtained with
this method coincide with the values $x_0$ of the zinc
concentration in the initial charge with precision better than $x_0~\pm~ 0.03$
for all samples with $x < x_c$.

The dependence of the crystal cell parameters $a$,$b$,$c$ of the single crystals of
\Li\ on zinc concentration $x$ in initial charge is shown in Fig.~\ref{fig:2}.
The solid symbols show the results obtained from the samples described above.
In contrast with the data of Ref.~\onlinecite{Hsu_2010} (open symbols),
we did not observe any anomaly in the concentration dependence of the cell
constants at $x=0.055$.

Magnetization curves in static magnetic fields of up to 7~T were measured with
a commercial SQUID magnetometer (Quantum Design MPMS-XL7).

The ESR experiments were performed with a transmission--type spectrometer using
resonators in the frequency range $18<\nu<120$~GHz. The magnetic field of a
superconducting solenoid was varied in the range of $0<\mu_0H<8$~T.
Temperatures were varied within the range of $1.2<T<30$~K.

The NMR experiments were performed with a conventional phase coherent,
homemade pulse spectrometer at a fixed frequency of  $\nu=61$~MHz. We investigated
the $^{7}$Li ($I=3/2$, $\gamma/2\pi=16.5466$ MHz/T) nuclei using spin--echo
technique with a pulse sequence 5$\mu$s--$\tau_D$--10$\mu$s, where the time
between pulses, $\tau_D$, was 40 $\mu$s. The spectra were collected by sweeping
the applied magnetic field between $3.5<\mu_0 H<3.9$~T. The
temperatures were stabilized with a precision better than 0.02~K.

\section{Experimental results}

\subsection{Magnetization measurements}

The experimental curves $M(T)/H$ and their temperature derivatives for
different field directions $\vect{H}\parallel \vect{a}, \vect{b}$ and
$\vect{H}\parallel\vect{c}$ at $\mu_{0}H=0.1$~T are shown in
Figs.~\ref{fig:M(T)} and \ref{fig:dmdt}. The dependences were obtained on the
single crystal without doping ($x=0$) and on the twinned single crystal with
$x=0.1$. For all field directions broad maxima of $M(T)/ H$ were observed at
$T=38$~K for samples without doping and at $T=31$~K for doped samples with
$x=0.1$. The maximum is typical for low-dimensional antiferromagnets. In the
paramagnetic region, the magnetic susceptibility for $\vect{H}\parallel
\vect{c}$ exceeds that for $\vect{H}
\parallel \vect{a}, \vect{b}$, which is consistent with the anisotropy of the
$g$-tensor measured in the ESR experiments~\cite{Vorotynov_1998} ($g_{a,b}=
2.0$; $g_{c}=2.2$). At high temperatures $T \gtrsim 150$ K the susceptibility
of the doped sample is $(20\pm1.2)\%$ less then that of the pure sample. This
reduction is close  to the factor $(1-2x)$ expected if all Zn$^{2+}$ ions take
the places of the Cu$^{2+}$ ions, which would result in the corresponding
decrease of the number of paramagnetic spins. To emphasize this fact the
$M(T)/H$ dependence for the sample without doping reduced by factor 0.8 is
shown in the same figure with dashed lines. This experiment allows to conclude
that  the number of magnetic Cu$^{2+}$ ions in the doped samples of \Li\ is
reduced by the number of nonmagnetic Zn$^{2+}$ ions.

The transitions to the magnetically ordered states at $T_{c1}$ and $T_{c2}$ are
marked by the change of the $M(T)/H$ slope. These inflection points are well
resolved in the temperature derivative of $M(T)/H$ for the undoped sample
(Fig.~\ref{fig:dmdt}). The $(d/dT)(M(T)/H)$ curves are strongly
anisotropic\cite{Bush_2012}: for $\vect{H} \parallel \vect{b}$ and $\vect{c}$,
there are two anomalies corresponding to two transitions at $T_{c1}$ and
$T_{c2}$. The temperature dependence of $(d/dT)(M(T)/H)$ for $\vect{H}\parallel
\vect{a}$ shows only one sharp peak at the lower temperature near $T_{c2}$ in
the studied field range. For the doped sample the inflection points on the
$M(T)/H$ curves were smoothed, but the maximum of the slope for $\vect{H}
\parallel \vect{c}$ was observed at evidently higher temperature than for
$\vect{H} \parallel \vect{a},\vect{b}$. The distance between the maxima for
doped samples with $x < 0.1$ are almost the same as for the samples without
doping (Fig.~\ref{fig:dmdt}). Thus the double-step transition in the
magnetically ordered phase also takes place for doped samples. Note here that
the double-stage transition is specific for a planar spin structure with a
strong easy-plane anisotropy for the vector $\vect{n}$ normal to the spin
plane. The anomaly of $M(H)$ at a field of $\mu_0 H_{c1}\approx 2$ T for
$\vect{H}\parallel \vect{b}$ which was associated with the reorientation of the
spin plane observed earlier for samples without doping \cite{Bush_2012}
 was not observed for all studied samples with $x \geq 0.025$.

\subsection{Electron Spin Resonance experiments}
\begin{figure}
\includegraphics[width=\figwidth,angle=0,clip]{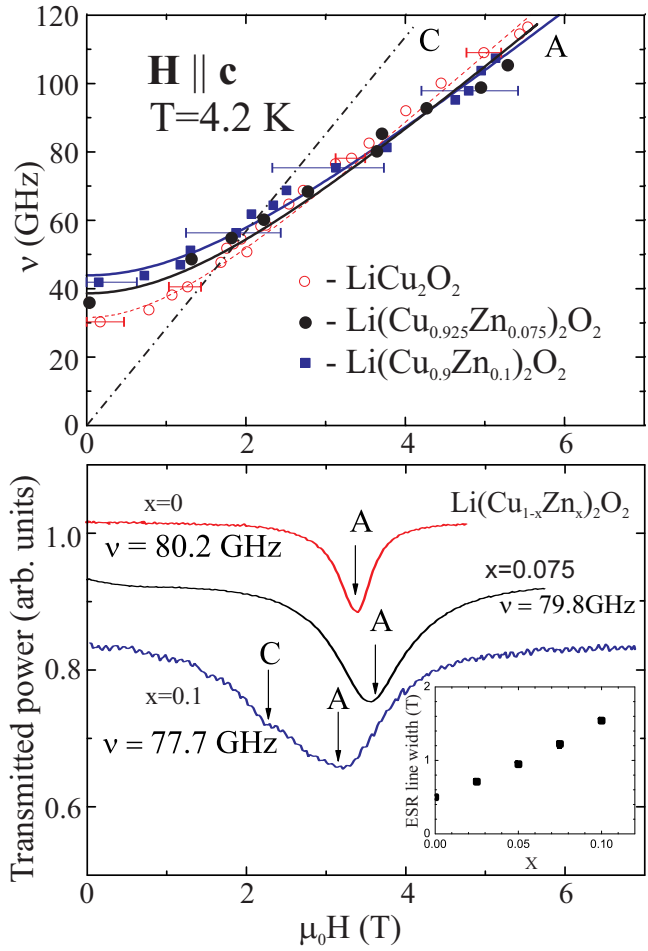}
\caption{(color online) Upper panel: ESR frequency-field dependences for \Li\
samples with  $x = 0.075, x=0.1$ (black and blue solid symbols), and $x=0$
(open symbols, from Ref.~\onlinecite{Svistov_2009}), $\vect{H}\parallel
\vect{c}$, $T=4.2$ K. Solid and dashed lines show fits with the equation given
in the text. The dash-dotted line shows the spectra of the absorption component
``C'', which coincides with the position of the paramagnetic absorption at
$T>T_N$. Lower panel: Examples of ESR absorption spectra measured on three
samples with $x=0.1$, $x=0.075$, and $x=0$ at close frequencies ($\nu\approx
79$ GHz), $\vect{H}\parallel \vect{c}$ and $T=4.2$ K. Insert to the lower
panel: dependence of the ``A''-component ESR line width on the Zn
concentration. $\vect{H}\parallel \vect{c}$, $T=4.2$~K, $\nu \approx 79$ GHz.
 }
\label{fig:4}
\end{figure}

\begin{figure}
\includegraphics[width=\figwidth,angle=0,clip]{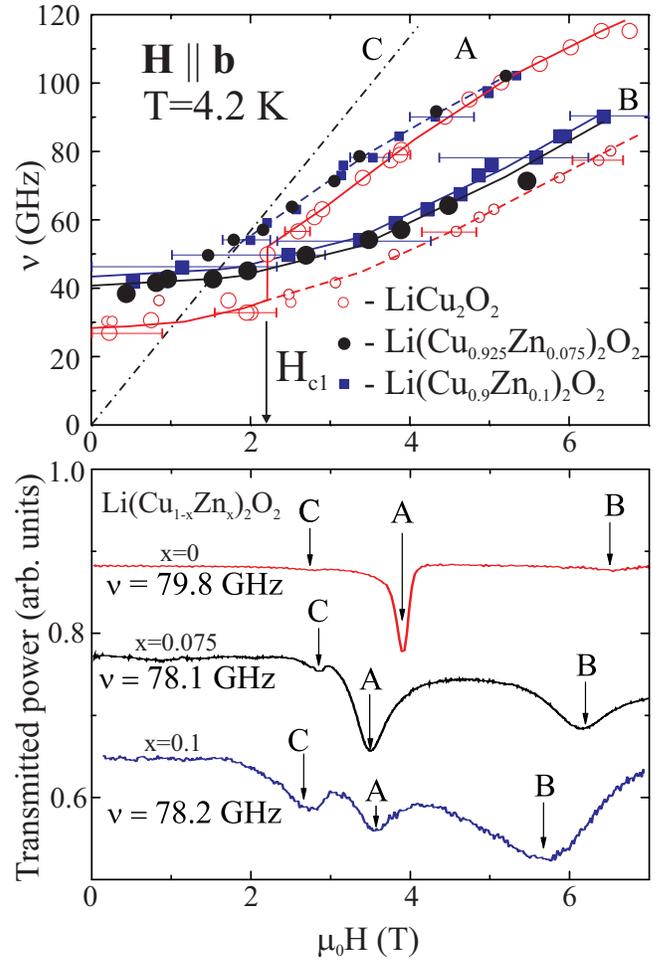}
\caption{(color online) Upper panel: ESR frequency-field dependences for \Li\
samples with  $x = 0.075, x=0.1$ (black and blue solid symbols), and $x = 0$
(open symbols, from Ref.~\onlinecite{Svistov_2009}). Intense components are
marked by large symbols, weak components are marked by
 small symbols, respectively, for $\vect{H}
\parallel \vect{b}$, $T = 4.2$ K. The solid and dashed lines are guides to the eyes. The
dash-dotted line corresponds to the absorption component ``C'', which coincides
with the EPR spectra measured at $T>T_N$.  Lower panel: Example of field
dependences of the transmitted power, measured on three samples with $x = 0.1$,
$x=0.075$, and $x = 0$ at two close frequencies ($\nu\approx 79$ GHz),
$\vect{H}\parallel \vect{b}$ and $T = 4.2$ K. } \label{fig:5}
\end{figure}

\begin{figure}
\includegraphics[width=\figwidth,angle=0,clip]{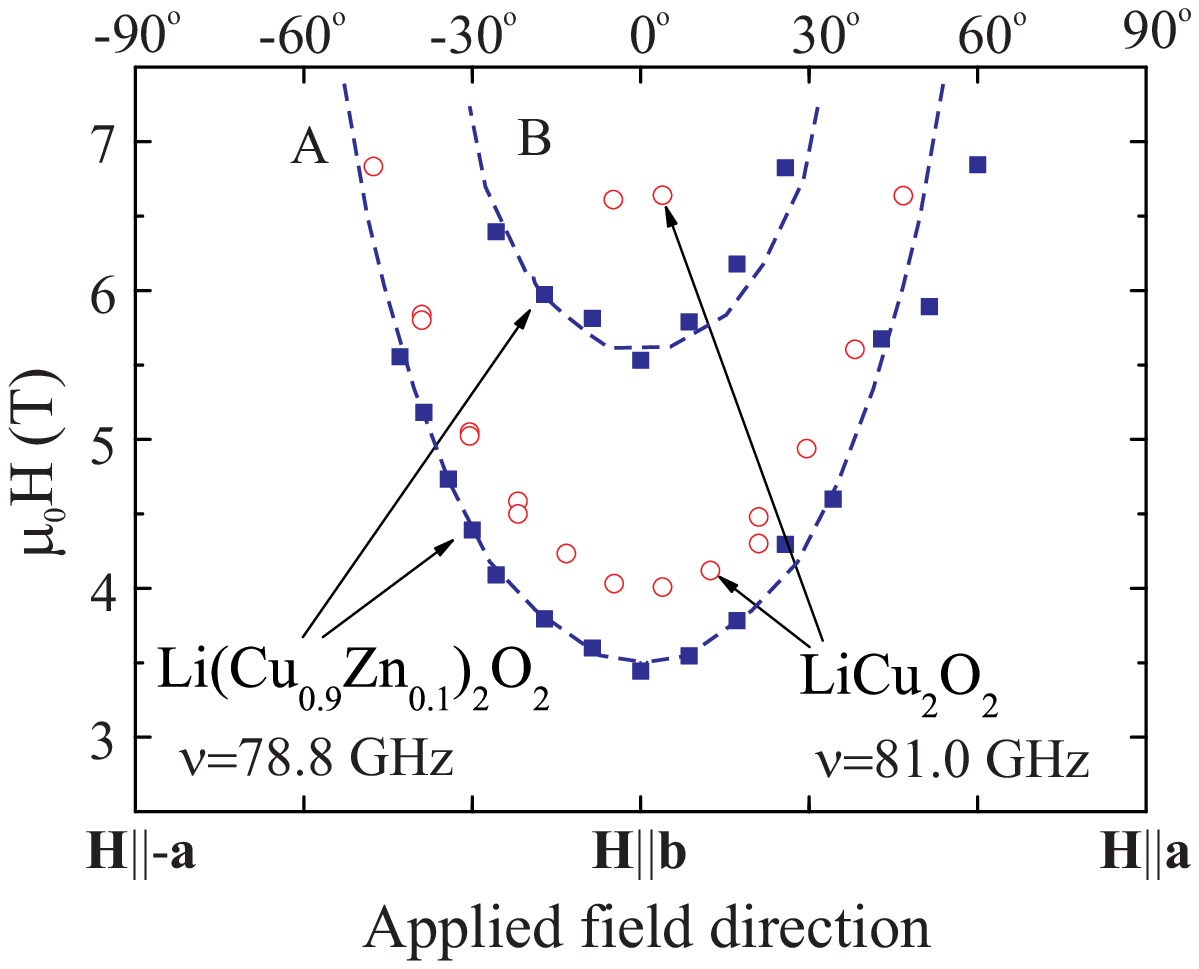}
\caption{(color online) ($ab$)-plane angular dependences of  the resonance
fields for ``A'' and ``B'' components measured on \Li\ samples with $x = 0.1$
(solid symbols), and $x = 0$ (open symbols) for frequencies around 80 GHz. $T =
4.2$ K. The dashed lines are guides to the eyes.} \label{fig:6}
\end{figure}

The ESR in single crystals of \Li\ was studied on the samples with $x = 0.025,
0.05, 0.075, 0.1$. The samples with $x = 0.025$ and $x=0.1$ were untwinned. ESR
on the sample without doping was studied in a previous work.\cite{Svistov_2009}
The typical ESR absorption lines are shown in the lower panels of
Figs.~\ref{fig:4} and \ref{fig:5}. We present the lines for the samples with
the  Zn concentration $x=0.075$, $x=0.1$ and the undoped untwinned sample from
Ref.~\onlinecite{Svistov_2009}. The ESR absorption lines are systematically
broader for the samples with larger Zn concentration. The insert in the lower
panel of Fig.~\ref{fig:4} shows the concentration dependence of the absorption
line width for $\vect{H}\parallel \vect{c}$ at $T=4.2$ K. The ESR
frequency-field dependences $\nu(H)$ for field directions $\vect{H}\parallel
\vect{c}$, $\vect{b}$ for the samples with $x=0$, $0.075$ and $0.1$ are shown
in the upper panels of Figs.~\ref{fig:4} and \ref{fig:5}. Note, that for
$\vect{H}\parallel \vect{c}$ the presence of the twinning is not essential. The
sample with $x=0.05, 0.075$ were twinned, but due to strong anisotropy (see
below) it was easy to separate the absorption lines only from the domain with
$\vect{H}\parallel\vect{b}$. The dash-dotted lines in the upper panels of
Figs.~\ref{fig:4} and \ref{fig:5} mark $\nu(H)$ dependences of electron
paramagnetic resonance for LiCu$_2$O$_2$: $\nu=(g\mu_B/h)H$ ($g_{a,b}= 2.0$;
$g_{c}=2.2$).\cite{Vorotynov_1998} Resonance fields of the relatively weak
component ``C'' are well described by such a paramagnetic dependence. Probably,
this component is due to paramagnetic inclusions.

 The experimental data which correspond to the absorption
component ``C'' are not shown in the upper panels of Figs.~\ref{fig:4} and
\ref{fig:5}. The ESR frequency-field dependences ($\nu (H)$) of the
intermediate Zn concentrations $x=0.025$, $x=0.05$ (not shown) and $x=0.075$
interpolate between those of the samples with extreme doping concentration
$x=0$ and $x=0.1$. The error bars in the upper panels of Figs.~\ref{fig:4} and
\ref{fig:5} show the antiferromagnetic resonance line widths measured at the
half of the absorbed intensities.

The main features of the ESR frequency-field dependences ($\nu (H)$) for
different Zn concentrations remain the same as for the undoped sample. The
frequency-field dependences ($\nu (H)$) consist of a single branch for
$\vect{H}\parallel \vect{c}$ and two branches for $\vect{H}\parallel \vect{b}$.
In the case of pure compound mode ``A'' was ascribed to the oscillations of the
spin plane around $\vect{a}$-axis.\cite{Svistov_2009} The sudden change of this
mode resonance frequency at $\vect{H}||\vect{b}$ at $H_{c1}$ is associated with
spin-flop transition.

The zero-field gap grows on doping  from $\nu(H=0)=30\pm 2$ GHz for the undoped
sample ($x=0$) to $\nu(H=0)=42 \pm 2$ GHz for doped samples with $x=0.1$.
Angular dependences of the resonance fields on rotation in the $(ab)$-plane for
samples with $x=0, 0.1$ are shown in Fig.~\ref{fig:6}. Both resonance fields
shift to higher fields as $\vect{H}$ approaches $\vect{H}\parallel \vect{a}$.
Such a behavior was observed also for a rotation of the applied magnetic field
$H$ in the $(ac)$-plane.  Thus, rotation of the static field towards the
$\vect{a}$-direction flattens the $\nu(H)$ dependence, which finally becomes
field independent for $\vect{H}\parallel \vect{a}$ (as was reported for the
pure compound in Ref.~\onlinecite{Svistov_2009}). From these data we conclude
that the strong uniaxial anisotropy along the $\vect{a}$-direction remains for
all studied doped samples. Such an anisotropy corresponds to the easy
$(bc)$-plane anisotropy for the vector $\vect{n}$ normal to the spin plane.

The branches marked as "A" for $\vect{H}\parallel\vect{c}$  are quasi linear
with field: $\nu=k\sqrt{H^2+\Delta^2}$. The coefficient $k$ is noticeably
smaller than the gyromagnetic ratio  $g\mu_B/h$.  Such a field dependence is
typical for a planar spin structure with strong ``easy-plane'' anisotropy for
vector $\vect{n}$ perpendicular to the spin plane and field direction
perpendicular to the anisotropy axis ($\vect{H}\perp\vect{a}$). (See attachment)
The coefficient $k$ is defined by the anisotropy of the susceptibility of the
spin structure: $k=(g\mu_B/h)\sqrt{(\chi_{\parallel}/\chi_{\perp}-1)}$ (here
$\chi_{\parallel}$ and $\chi_{\perp}$ are the susceptibilities for field
directions parallel and perpendicular to the vector $\vect{n}$ of the spiral
spin structure). Using the value of $k$ as a fit parameter we obtained that the
anisotropy of the susceptibility $\chi_{\parallel}/\chi_{\perp}-1$ slightly
decreases from $0.55\pm 0.02$ to $0.5\pm 0.02$  with the increase of doping
from $x=0$ to $x=0.1$.

For the field orientation $\vect{H}\parallel \vect{b}$ the magnetic structure
of the undoped LiCu$_2$O$_2$ samples undergoes a reorientation at a field of
$\mu_0 H_{c1}\approx 2$ T. Earlier experiments  have shown that at this field
the spin plane of the spiral structure rotates from the $(ab)$ to the
$(ac)$-plane.\cite{Svistov_2009,Bush_2012} The magnetic susceptibility at this
field undergoes a step like increase. The ESR frequency-field dependence for
this orientation is shown in the upper panel of Fig.~\ref{fig:5}. The
transition field $H_{c1}$ is marked by the jump of the frequency-field
dependence  for the undoped sample. The intensive component of the ESR
absorption line changes form the ``B''-branch of the frequency-field diagram to
the ``A''-branch at the transition field $H_{c1}$ for the undoped sample. Thus,
the ``A''-component dominates the absorption spectrum of the undoped sample at
high fields (Fig.~\ref{fig:5}). For the doped sample with $x=0.1$ the
absorption component ``B'' was more than 10 times more intensive than the
component ``A'' in the whole field range. No jumps of the position of the
intense component in the frequency-field diagram was observed for the doped
samples.

From these observations we conclude that for doped samples no magnetic
reorientation takes place in the  studied field range. This conclusion is in
agreement with the absence of anomalies in the $M(H)$ curves mentioned in the
previous subsection. Note that the value of $H_{c1}$ and the sharpness of the
transition are strongly dependent on the quality of the samples even for the
crystals without doping.\cite{Svistov_2009, Bush_2012}

\subsection{Nuclear Magnetic Resonance experiments}
\begin{figure}
\includegraphics[width=\figwidth,angle=0,clip]{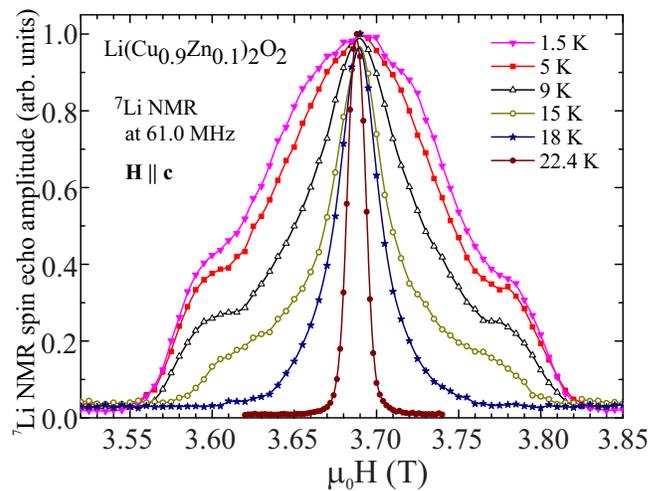}
\caption{(color online) NMR spectra of $^7$Li nuclei in \Li, $x=0.1$, measured at
different temperatures. $\vect{H}\parallel \vect{c}$ and pumping frequency
$\nu=61.0$ MHz.
 }
\label{fig:7}
\end{figure}

\begin{figure}[!ht]
\includegraphics[width=\figwidth,angle=0,clip]{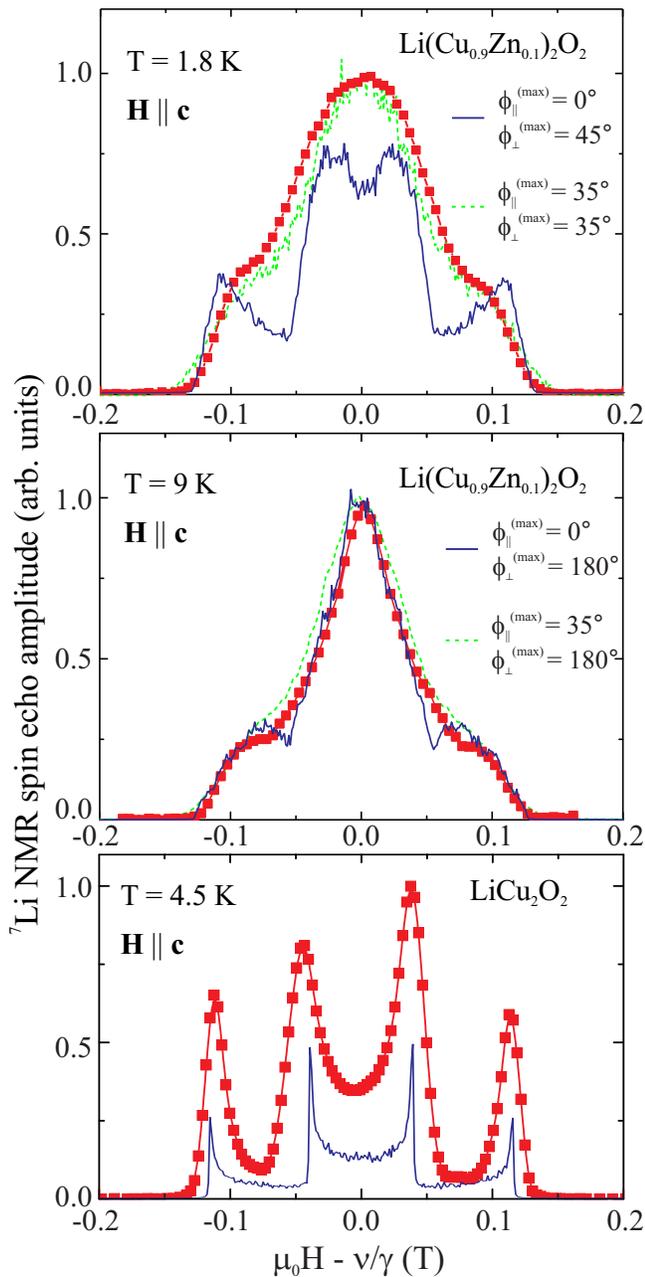}
\caption{ (color online) Squares represent NMR spectra of $^7$Li nuclei in \Li,
$x=0.1$ (top and middle panels) and $x=0$ (bottom panel) measured at $T < T_N
$ for $\vect{H}\parallel \vect{c}$ and $\nu=61.0$ MHz. The solid lines show the
simulated spectra in models of spiral planar spin structures within the
$(ab)$-plane with long-range order in the $\vect{c}$-direction for the undoped
sample (bottom panel) and with disorder in the $\vect{c}$-direction for the
doped sample (top and middle panel). The dashed lines show the simulated
spectra in the models of randomly disturbed planar spiral spin structure within
$ab$-planes with short-range order in $c$-direction (top and middle panels).}
\label{fig:8}
\end{figure}

The magnetic structure of the single crystals of \Li\ with $x=0.1$ was tested
with the NMR technique. $^7$Li NMR field-sweep spectra measured at different
temperatures at field orientation $\vect{H}\parallel \vect{c}$ and fixed
frequency $\nu=61.0$ MHz are shown in Fig.~\ref{fig:7}. The shape of the NMR
spectra transforms from a single component spectrum at high temperatures
$T>T_N$ to a spectrum with characteristic shoulders at temperatures below
$T_N\approx 18$ K (see Fig.~\ref{fig:7}). Similar NMR spectra evolution was
reported in Ref.~\onlinecite{Hsu_2010-1} for the sample with Zn content
$x=0.029$.

The spin echo NMR signal in the low temperature range was observed in a broad
field range $\mu_0\Delta H \approx 0.25$ T which is approximately the same as
for LiCu$_2$O$_2$ without doping (see Fig.~\ref{fig:8}).  At temperatures close
to $T_N$ the central line of the NMR spectra dominates. This fact demonstrates,
that close to Neel temperature the number of nuclei with paramagnetic
environment increases.

In Fig.~\ref{fig:8} the lines show the fitting results of the NMR spectra. The
$^7$Li NMR spectra for doped and undoped samples were fitted by taking into
account dipolar fields within the model of a planar spiral magnetic structure
in the $(ab)$-plane  with an incommensurate wave vector $k_{ic}=0.827 \times 2
\pi /b$ aligned along the $\vect{b}$-direction and an effective magnetic moment
on the Cu$^{2+}$ position equal to $0.8 \mu_B$. Note, that the contribution of
the contact field for such spin arrangement at $\vect{H}\parallel\vect{c}$ can
be neglected since the spin plane is perpendicular to the magnetic field.\cite{chemshift} The magnetic structure for the sample without doping is
described by Eqn.(1). The mutual orientation of the spins from neighboring
$(ab)$-planes defined by phases $\phi_{\alpha, \beta,\gamma,\delta}$ are given
in the end of the section II. The line width of individual groups of resonating
lithium nuclei for this computation was taken as $\delta H=0.005$ T, which is
much smaller than all peculiarities on the experimental spectra. This value of
$\delta H$ roughly corresponds to the value of quadrupolar splitting of NMR
line on $^7$Li nuclei in LiCu$_2$O$_2$ ($\nu_q\approx 51.7$ kHz,
Ref.~\onlinecite{Gippius_2004}).

To explain the main features of our experimentally observed NMR spectra we
consider effective fields at lithium sites which are generated by an individual
magnetic chain of Cu$^{2+}$ ions nearest to the lithium nuclei (see the
fragment of the spiral structure shown in Fig.~\ref{fig:spins}c). For
simplicity we suppose that the Li$^+$ ion, the Cu$^{2+}$ chain, the spin
vectors, and the applied static field $\vect{H}$ lies in the plane of the
figure. The projection of the effective field from neighboring spins on the
applied field $\vect{H}$ changes along the spin chain sinusoidally. For a given
propagation vector $k_{ic}$ the NMR spectra exhibit a shape with two
characteristic maxima at the extremum fields $\nu/\gamma  \pm H_{extr}$. Since
the Li$^+$ ions are located symmetrically between two Cu$^{2+}$ ions, the
extreme fields $H_{extr}$ will correspond to the arrangement of the spins in
the copper chain symmetric with respect to the chosen Li$^{+}$ ion (marked with
index $i$ in Fig.~\ref{fig:spins}c). As it is schematically shown in
Fig.~\ref{fig:spins}c, such a symmetric arrangement corresponding to the
magnetic structures, with opposite wave vectors $k_{ic}$ (solid arrows) and
$-k_{ic}$ (dotted arrows), yields different dipolar contributions to effective
field. Thus for such simple model structure we can expect a broad NMR spectrum
with four characteristic maxima corresponding to two magnetic domains with
opposite wave vectors.

The magnetic structure expected for LiCu$_2$O$_2$ at an arbitrary orientation
of the applied magnetic field provides NMR spectra with 8 maxima corresponding
to the four extremal fields at the lithium nuclei from different positions Li1,
Li2, Li3, and Li4 (Fig.~\ref{fig:spins}b). For the field orientation
$\vect{H}\parallel \vect{c}$ the number of maxima of the spectra is reduced to
four (see bottom panel of Fig.~\ref{fig:8}).

 In order to interpret the NMR spectra of doped samples we
simulated the NMR spectra for the structures with random static deviations of
the spin directions from the spin orientations of undoped model. For modeling
purposes the deviation angles within the spin plane were set to the random
value $-\phi^{(max)}<\phi<\phi^{(max)}$, where the maximal deviation
$\phi^{(max)}$ was used as a model fitting parameter. Taking into account that
the exchange interactions between planes is weaker than the in-plane
interactions, we used different deviation parameters $\phi^{(max)}_\parallel$
for spins within $(ab)$-plane, and $\phi^{(max)}_\perp$ for spins from
different planes. The NMR spectra of doped samples at 9 K can be fitted in the
model of totally random mutual orientation of the spins of the neighbor planes,
i.e. $\phi^{(max)}_\perp=180^\circ$ (see middle panel of Fig.~\ref{fig:8}). The
NMR spectra obtained in this model are only weakly sensitive to the disorder
within the individual spin plane. The solid blue line shows the computed NMR
spectra with the assumption of ordered moments within each $(ab)$-plane
($\phi^{(max)}_\parallel=0^\circ$). The dashed green line shows NMR spectra
obtained under the assumption that the spin directions of ions within the
$(ab)$-planes are oriented randomly around undisturbed directions within the
angle $\phi^{(max)}_\parallel=35^\circ$. For the fit we took $0.8\mu_B$ as the
absolute value of the magnetic Cu$^{2+}$ moments at this particular temperature
of $T = 9$ K. As temperature decreases the central maximum of the spectra
starts to broaden (see Fig.~\ref{fig:7}). Such transformation can be explained
by the development of spin correlations between ions of neighboring
$(ab)$-planes on cooling. The NMR spectrum measured at $1.8$ K (squares) and
the fitting results (lines) are given in the top panel of Fig.~\ref{fig:8}.
Again, the solid blue line shows the computed NMR spectrum with the assumption
of ordered moments within each (ab)-plane ($\phi^{(max)}_\parallel=0^\circ$)
and random deviations of nearest spins from neighboring planes within the angle
$\phi^{(max)}_\perp=45^\circ$. A better agreement between our model and the
low-temperature experimental NMR spectrum is obtained for the case of a
magnetic structure with short ranged static correlations for both in-plane and
inter-plane correlations according to the green dashed line. It shows the
computed NMR spectrum with the assumption of random deviations of spins
described by the parameters
$\phi^{(max)}_\parallel=\phi^{(max)}_\perp=35^\circ$. Again, the absolute value
of the magnetic moments of Cu$^{2+}$ was taken to be $0.8\mu_B$ for the fit.

Finally, we must note that the local field at the probing nuclei is defined
mostly by four nearest coordination spheres, which means that the NMR spectra
are testing the short-range static correlations only. These correlations must
be static at least during the time window of $\approx 0.1$ s set by our NMR
experiment. Thus, the simulation  does not actually rely on the assumption of
the long-range order and the same results can be obtained if we assume presence
of  only short-range order superimposed by the static random variations.

\section{Discussion}

X-ray diffraction and electron probe micro analysis of the single-phase
crystals of \Li\ show that Zn ions enter into the lattice without destruction
of the crystal structure in the broad concentration range $0\leq x \lesssim
0.12$. The scaling of the high-temperature susceptibility as $(1-2x)$, shown in
the Fig.~\ref{fig:2}, indicates that Zn ions substitute for Cu$^{2+}$ and each
Zn$^{2+}$ ion reduces the number of magnetic ions by one. Nonmagnetic dilution
should also affect the effective interspin interactions (i.e. Curie-Weiss
temperature $\Theta$). The fit of our $H||c$ susceptibility data in the $150
K-300 K$ temperature range shows that the Curie-Weiss temperature decreases
from $(61\pm3)K$ for the pure compound to $(51\pm5)K$ for the $10\%$ doped
compound. The Curie-Weiss temperature of the doped system is also about $80\%$
of that for the pure compound, which is again close to the $(1-2x)$ factor
describing the decrease of the average number of neighbors of the magnetic ion
with doping. On the other hand, even for the $10\%$ doping the $M(T)$ curves
demonstrate the same characteristic broad maximum around $35K$ as for the pure
compound. Thus, magnetization measurements prove that the magnetic subsystem of
\Li\ can still be considered as a system of spins $(S = 1/2)$ with frustrated
exchange interactions, where the spins are arranged in chains along the
crystallographic b-axis.

According to Ref.~\onlinecite{Masuda_2005} the strongest exchange interactions
for undoped \Li\ are the interactions between copper ions within the
($ab$)-planes: intrachain interactions of nearest spins  $J_1$,  intrachain
interactions of next-nearest spins $J_2$ and interchain interaction $J_3$ (see
Fig.~\ref{fig:spins}). These interactions lead to a spiral long-range magnetic
order with an incommensurate wave vector directed along the crystallographic
$\vect{b}$-axis. The solitary nonmagnetic Zn$^{2+}$ defects disturb this spiral
magnetic structure in their immediate surroundings only. Since a nonmagnetic
defect breaks up nearest-neighbor exchange bonds along the chain, but leaves
next-nearest exchange bond almost unperturbed, the effect of the solitary
defect on the spiral structure can be envisioned as a spiral phaseshift at the
defect location. Interchain couplings fix relative phases of the neighboring
spirals far away from the solitary defect which results in the formation of the
disturbed area of the finite size. For small concentrations $x$ the
characteristic size of the disturbed area around nonmagnetic defects can be
evaluated as $\frac{2(J_1,J_2)}{J_3}$ cell units, which yields a value of 4
cell units in the case of LiCu$_2$O$_2$. For $x=0.1$ the average distance
between the defects within the chains amounts to $\approx 5$ lattice constants
and, therefore, the disturbed areas strongly overlap.

As the defect positions in the neighboring chains are uncorrelated, the
interchain coupling became frustrated since all interchain bonds obviously can
not be satisfied simultaneously because of the spiral phaseshift at the defect
location. The doped frustrated quasi-one dimensional system differs
significantly from the non-frustrated doped system: in the case of the
non-frustrated system fragments of the spin chain on the both sides of the
defect are decoupled and weak interchain interactions can be completely
satisfied by properly setting orientation of the each fragment. Thus, at the
strong doping level long-range order in the frustrated quasi-one dimensional
system is most likely destroyed, while the short-range chiral correlations
should persist. We suggest that such a disordered state of the frustrated spin
chain can be a realization of a novel spin-glass like state with short-range
static chiral correlations in the $(ab)$-planes. The possibility of such state
in Zn doped LiCu$_2$O$_2$ was suggested earlier by Hsu at al.~\cite{Hsu_2010-1}

The ESR experiments show that the main features of the low-frequency spectrum
of excitations in \Li\ do not change at a doping of $x \leq 0.1$. The spectra
can be explained in the frame of a planar spin structure with strong easy-axis
anisotropy along the a-axis of the crystal. This means that the spins of this
spin-glass like structure lie in the plane defined by the applied magnetic
field $\vect{H}$ and the axial-anisotropy vector along the a-axis for all
concentrations $x$ within the entire range $0 \leq  x < 0.1$. The absorption
line width of \Li\ strongly increases with the Zn concentration (insert in
Fig.~\ref{fig:4}) and the absorption lines of the highly doped samples are
asymmetric: the low-field part of the absorption lines is more broadened than
the high-field part. Probably, this asymmetry is due to a nonuniformity of the
magnetic structure of the doped \Li. The defects in the magnetic structure
might trigger excitations of magnons with nonzero wave vectors (see for example
Ref.~\onlinecite{Krug_2010}). The resonance condition for such excitations is
expected for applied magnetic fields lower than the antiferromagnetic resonance
field.

The NMR experiments indicate that the value of the magnetic moments at the
position of Cu$^{2+}$ ions at low temperature in doped samples is nearly the
same as in the samples without doping at temperatures less then 15 K.
Additionally, the shape of the NMR spectra of zinc doped samples indicates
short-range static correlations the same as in undoped samples. These
correlations appear at the temperature $\approx18$ K, where the
anomaly on magnetization curve was observed and the ESR spectra start to be gaped. The
temperature evolution of the shape of NMR spectra of doped samples can be
explained by appearance of short-range correlations between spins of neighbor
$(ab)$-planes with decrease of the temperature from 9 to 1.8 K. The
interpretation of the experimental spectra and our fitting results is natural
and one of the simplest. But we can not exclude other models, which can
describe the NMR spectra in another way. The results of NMR experiments on
strongly doped \Li, for $x = 0.1$ are in agreement with the spin-glass like
magnetic structure suggested above.

\section{Conclusions}
Untwinned single-crystalline samples of \Li\ were grown for $0\leq x<0.12$. It is
shown that the zinc doping diminishes the number of magnetic copper ions
Cu$^{2+}$ and a single crystallographic phase is maintained for all samples
within the entire doping range $0\leq x<0.12$. The ESR spectra for all doping
concentrations within $0 \leq  x < 0.12$ can be well explained in the model of
a planar spin structure with strong easy-plane type anisotropy for the vector
$\vect{n}$ normal to the spin plane. The NMR spectra of the highly doped single
crystal Li(Cu$_{0.9}$Zn$_{0.1}$)$_2$O$_2$ can be well described by planar
spin-glass like magnetic structure with static short-range spiral correlations. The value
of magnetic moments of Cu$^{2+}$ ions in this structure is close to the value
obtained for undoped crystals: (0.8 $\pm$ 0.1) $\mu_B$.

\section{Attachment}

According to the theory of exchange symmetry~\cite{Andreev_1980} the order parameter in a coplanar exchange structure is represented by two mutually perpendicular unit spin vectors $\mathbf{l_1}$ and $\mathbf{l_2},$ which transform according to a particular irreducible  representation of the crystal symmetry group. In a helical magnetic structure, the order parameter with respect to the subgroup of translations transforms according to the representation with an incommensurate wave vector. Lagrange function for this system can be written as
$$L=\frac{I_1}{2}((\mathbf{\dot l_1}+\gamma[\mathbf{l_1 H}])^2+(\mathbf{\dot l_2}+\gamma[\mathbf{l_2 H}])^2)+\frac{I_2}{2}(\mathbf{\dot n}+\gamma[\mathbf{n H}])^2-U_a,$$

where $\mathbf{n} = [\mathbf{l_1 l_2}]$, and $U_a$ - small addition to the exchange part of the energy due to relativistic effects. $\mathbf{I_i}$ and the susceptibility tensor components are related by the expressions: $\chi_\perp= I_1 + I_2,\chi_\parallel  = 2 I_1$. In the spiral magnetic structure, $U_a$ depends only on the orientation of the vector $\mathbf{n}$ with respect to the crystal axes. In the simplest case, when the direction of all the spirals coincide, senior terms  in the expansion of the quantity $U_a$ in the components of the vector $n$ for a biaxial crystal can be written as $\frac{1}{2}(An_z^2 + Bn_y^2)$. For a magnetic field directed along one of the symmetry axes, vector $\mathbf{n}$ also directed along one of the twofold axes. The coordinate axes are chosen so that, in a zero magnetic field, the vector $\mathbf{n}$ is parallel to the $\mathbf{z}$-axis. Then, the coefficients $A$ and $B$ satisfy the inequalities $A < 0$, $A < B$. In the absence of the magnetic field, oscillation frequencies are
$$\omega_{10}^2 = \frac{-A}{\chi_\perp},~ \omega_{20}^2 = \frac{B-A}{\chi_\perp},~ \omega_{30} = 0$$

\begin{figure}
\includegraphics[width=\figwidth,angle=0,clip]{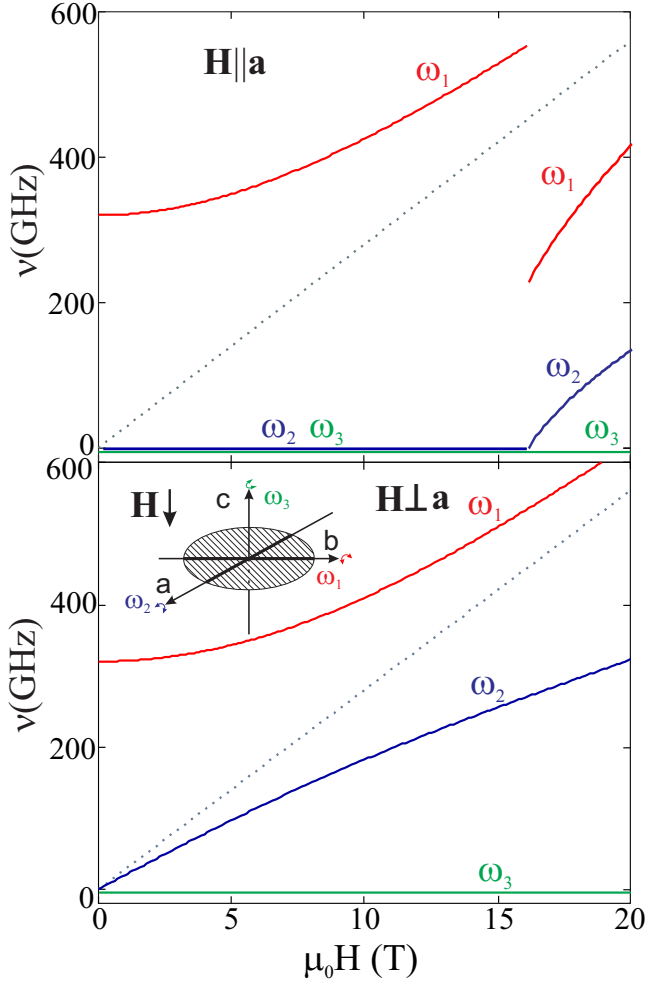}
\caption{(color online) The calculated ESR spectra of the planar spiral antiferromagnet with uniaxial anisotropy and parameters of LiCu$_2$O$_2$ ($\omega_{10} = 320$ GHz, $\omega_{20}=0 $ GHz and $\sqrt{\frac{\chi_\perp}{\chi_\parallel - \chi_\perp}} = 1.4$). Solid lines show the spectra. Top panel: $\vect{H} || \vect{a}$. Bottom panel: $\vect{H}\perp\vect{a}$. Dotted line indicate the paramagnetic resonance spectra. Insert to the bottom panel: the scheme of oscillations of the spin plane at $\vect{H} || \vect{c}$.}
\label{fig:1}
\end{figure}

The third oscillation frequency is equal to zero also for $H \ne 0$. In the magnetic field directed along the $\mathbf z$-axis, the resonance frequencies are determined by the formula
\begin{multline}
\omega_{1,2}^2 = \frac{\omega_{10}^2 + \omega_{20}^2}{2}+\gamma^2H^2\frac{I_1^2 + I_2^2}{(I_1 + I_2)^2} \pm \\ \pm \left( \frac{(\omega_{10}^2 - \omega_{20}^2)^2}{4} + 2(\omega_{10}^2 + \omega_{20}^2)\gamma^2H^2\frac{I_2^2}{(I_1 + I_2)^2}  \right. + \\ + \left. 4\gamma^4H^4\frac{I_1^2I_2^2}{(I_1 + I_2)^4}\right)^{1/2}.
\end{multline}

For $\mathbf {H \parallel y}$
$$\omega_1^2 = \omega_{10}^2,~ \omega_2^2 = \omega_{20}^2 + \gamma^2 H^2$$

For $\mathbf {H \parallel x}$
$$\omega_1^2 = \omega_{10}^2 + \gamma^2 H^2,~ \omega_2^2 = \omega_{20}^2$$

The reorientation transitions occurs in the following fields:

If $\mathbf {\chi_\perp > \chi_\parallel}$ and $\mathbf{H \parallel z}$
$$H_{cz} = \frac{min(\omega_{10},\omega_{20})}{\gamma} \sqrt{\frac{\chi_\perp}{\chi_\perp - \chi_\parallel}}$$

And if $\mathbf {\chi_\perp < \chi_\parallel}$ and $\mathbf H$ directed along $\mathbf x$ or $\mathbf y$
$$H_{cx} = \frac{\omega_{10}}{\gamma} \sqrt{\frac{\chi_\perp}{\chi_\parallel - \chi_\perp}},~H_{cy} = \frac{\omega_{20}}{\gamma} \sqrt{\frac{\chi_\perp}{\chi_\parallel - \chi_\perp}},$$
respectively. In the first case, the oscillation frequencies in fields above the field of the spin-flop are given by
$$\omega_1^2 = \frac{A}{\chi_\perp} + \gamma^2 H^2,~\omega_2^2 = \frac{B}{\chi_\perp},$$
$$\omega_1^2 = \frac{-B}{\chi_\perp} + \gamma^2 H^2,~\omega_2^2 = \frac{A - B}{\chi_\perp},$$
depending on whether the constant B is positive $(\mathbf{n \parallel x}$ at $H > H_{cz})$ or negative $(\mathbf{n \parallel y}$ in $H> H_{cz})$. In the second case, the frequencies are determined by expression (1), into which is necessary to substitute
$$\omega_{10}^2 = \frac{A}{\chi_\perp},~\omega_{20}^2 = \frac{B}{\chi_\perp}$$
for $\mathbf{H \parallel x},~ H > H_{cx}$ and
$$\omega_{10}^2 = \frac{-B}{\chi_\perp},~\omega_{20}^2 = \frac{A - B}{\chi_\perp}$$
for $\mathbf{H \parallel y},~ H > H_{cy}$

In a more complex case, where not all the spirals twisted to one side, in the expansion of $U_a$ can also add member $Cn_y$. In this case, as well as the fields are not directed along the axes of symmetry of the crystal, expressions for the frequencies are too bulky and the orientation of the order parameter and the frequency of small oscillations are found by numerical calculations of the equilibrium orientation of the order parameter, relative to the crystal axes and the frequencies of small oscillations.

Fig.~\ref{fig:1} shows the calculated spectra for LiCu$_2$O$_2$ in the model of planar spiral antiferromagnet with uniaxial anisotropy: $A=B<0$ and parameters: $\omega_{10} = 320$ GHz, $\omega_{20}=0 $ GHz and $\sqrt{\frac{\chi_\perp}{\chi_\parallel - \chi_\perp}} = 1.4$.

\acknowledgements
We are thankful for the useful and enlightening discussions to S. S. Sosin. This
work is supported by the Grants 12-02-00557-a, 10-02-01105-a, 11-02-92707-IND-a
of the Russian Foundation for Basic Research, Program of Russian Scientific
Schools, and by the German Research Society (DFG) within the Transregional
Collaborative Research Center (TRR 80).


\begin{thebibliography}{30}
\bibitem{chemshift} The value of the chemical shift, as determined in our experiments at the paramagnetic phase (T=29K),
is equal to $0.075\pm0.005$ T/$\mu_B$ at $\vect{H}||\vect{c}$.
\end{thebibliography}
\end{document}